\documentclass[preprint2]{aastex}

\usepackage{threeparttable}

\slugcomment{}
\shorttitle{5.4 keV line in NGC\,4051}
\shortauthors{Turner et al.}
\def\suzaku{{\em Suzaku}\ }

\begin{document}

\title{{Significant X-ray Line Emission in the 5-6 keV band of NGC 4051}}
\author{T.J.Turner}
\affil{Department of Physics, University of Maryland Baltimore County, 
   Baltimore, MD 21250 and Astrophysics Science Division,   
NASA/GSFC, Greenbelt, MD 20771, U.S.A}

\author{L.Miller} 
\affil{Dept. of Physics, University of Oxford, 
Denys Wilkinson Building, Keble Road, Oxford OX1 3RH, U.K.}

\author{J.N.Reeves, A. Lobban}
\affil{Astrophysics Group, School of Physical and Geographical Sciences, Keele 
University, Keele, Staffordshire ST5 5BG, U.K}

\author{V.Braito}
\affil{Department of Physics and Astronomy, University of Leicester, 
University Road,  
Leicester LE1 7RH, UK}

\author{S.B.Kraemer}
\affil{Institute for Astrophysics and Computational Sciences, Department of Physics, 
The Catholic University of America, 
Washington, DC 20064; and Astrophysics Science Division, NASA Goddard Space Flight Center, 
Greenbelt, MD 20771}

\and 

\author{D.M.Crenshaw}
\affil{Department of Physics and Astronomy, Georgia State University, Astronomy Offices, One Park 
Place South SE, Suite 700, Atlanta, GA 30303, USA}

\begin{abstract}

A \suzaku X-ray observation of NGC 4051 taken during 2005 Nov reveals
line emission at 5.44\,keV in the rest-frame of the
galaxy which does not have an obvious origin in known rest-frame atomic
transitions.  The improvement to the fit statistic when this line 
is accounted for establishes its reality at $>99.9\%$ confidence:
we have also verified that the line is detected in the three XIS
units independently. Comparison between the data and Monte Carlo
simulations shows that the probability of the line being a statistical
fluctuation is $p < 3.3 \times 10^{-4}$. 
Consideration of three independent line detections 
in {\it Suzaku} data taken at different epochs yields a probability  
$p< 3 \times 10^{-11}$ and thus 
conclusively demonstrates that it cannot be a statistical fluctuation in the data.    
The new line and a strong component of Fe K$\alpha$ emission from neutral
material are prominent when the source flux is low, during 2005.
Spectra from 2008 show evidence for a line consistent with having the
same flux and energy as that observed during 2005, but inconsistent with having a
constant equivalent width against the observed continuum.  The
stability of the line flux and energy suggests that it may not
arise in transient hotspots, as has been suggested for similar lines
in other sources, but could arise from a special location in the
reprocessor, such as the inner edge of the accretion disk.
Alternatively, the line energy may be explained by spallation of Fe
into Cr, as discussed in a companion paper.

\end{abstract}


\section{Introduction}
New X-ray data from high-throughput missions such as
{\it XMM-Newton} and {\it Suzaku}, along with the high spectral 
resolution afforded by the {\it Chandra} 
High Energy Transmission Grating
(HETG) has led to the discovery of several
important spectral signatures in Active Galactic Nuclei (AGN) that
were not detectable using previous missions.

A joint {\it XMM-Newton}/{\it Chandra} observation of NGC 3516
\citep{turner02a} revealed the first detection of narrow line emission
at 5.6 and 6.2\,keV (the latter was able to be resolved from the
strong Fe K$\alpha$ line owing to the relatively high spectral
resolution afforded by the HEG grating).  While the line energies
matched those expected for ionized species of Cr and Mn, the line
strengths exceeded what might be expected from illumination of
material with cosmic abundance ratios.  One explanation considered was
enhancement of Cr and Mn abundances from spallation of Fe. However,
this was initially disfavored as the line ratios did not agree well
with those predicted by \citet{skibo97} for spallation of disk gas
\citep{turner02a}.  The alternative possibility, of Doppler-shifted
components of Fe emission, led to a suggestion of emitting hotspots on
the accretion disk surface - with an expectation of observable shifts
in line flux and energy as the hotspot traverses its orbit.  Further
examples of the phenomenon, dubbed `transient Fe lines', soon came
from observations of other AGN, with many lines reported in the
5-6\,keV regime \citep[e.g.][]{yaqoob03a,turner04a}.

In this paper we report the highly significant detection of an emission
line at 5.44\,keV  in the rest-frame of the nearby  
narrow line Seyfert 1 type AGN 
NGC\,4051. The systemic redshift of the host galaxy is 
$z=0.0023$ that yields a 
distance  9.3\,Mpc for $H_0=\,74$\,km\,s$^{-1}$\,Mpc$^{-1}$ assuming 
the redshift 
to arise from the Hubble flow.  However, for such nearby galaxies 
a more reliable estimation of distance can be obtained 
from use of the Tully-Fisher relation, giving 
a distance of 15.2\,Mpc \citep{russell04a} in this case, 
which we adopt in this paper. 
  
NGC\,4051 was observed by \suzaku in 2005 and 2008.  The 2005 observations
have previously been described and analyzed by \citet{terashima09a}, who showed 
that the X-ray source was highly variable on short timescales, and that 
a spectral model including variable 
partial-covering absorption was required to fit the data.
Here we make a joint analysis of the 2005 observation together with the new 
2008 data in which we concentrate specifically on the detection of the narrow emission
line and its variability.  We then discuss the possible origin  of the line in the 
context of  several popular models.

\section{Observations}

The \suzaku X-ray Imaging Spectrometer \citep[XIS][]{koyama07}
instrument comprises four X-ray telescopes \citep{mitsuda07} each
with a CCD in the focal plane.  XIS CCDs 0,2,3 are configured to be
front-illuminated (FI) and provide useful data over $\sim 0.6-10.0$\,keV
with energy resolution FWHM $\sim 150\,$eV at 6\,keV. XIS 1 is a
back-illuminated CCD and has an enhanced soft-band response (down to
0.2\,keV) but lower area at 6\,keV than the FI CCDs as well as a larger
background level at high energies, consequently this detector was not used in our analysis.
\suzaku also carries a non-imaging, collimated Hard X-ray Detector \citep[HXD][]{takahashi07}
whose PIN instrument provides useful data over 15-70\,keV for bright AGN. 

Our analysis used all available {\it Suzaku} observations of NGC 4051 
comprising data from 2005 Nov 10-13 (observation identifier 
700004010) and 2008 Nov 6-12 (703023010) and 23-25 (703023020) 
as summarized in Table~1. 
The data were reduced using v6.4.1 of {\sc HEAsoft} and screened to
 exclude: i) periods during and within 500 seconds of the South Atlantic
 Anomaly (SAA), ii) with an Earth elevation angle less than 10$^\circ$ and
 iii) with cut-off rigidity $>6$ GeV. The source was observed at the nominal
 center position for the XIS during 2008 and at the nominal center position for the 
HXD during 2005. The FI CCDs were in $3 \times 3$ and $5
 \times 5$ edit-modes, with normal clocking mode.  For the XIS we
 selected events with grades 0,2,3,4, and 6 and removed hot and
 flickering pixels using the SISCLEAN script.  The spaced-row charge
 injection (SCI) was used.  The XIS products were extracted from
 circular regions of 2.9\arcmin \, radius with background spectra from
 a region of the same size, offset from the source (avoiding the
 calibration sources at the edges of the chips). The response and
 ancillary response files were created using {\sc xisrmfgen v2007 May}
 and {\sc xissimarfgen v2008 Mar}.

NGC 4051 is too faint to be detected in the HXD GSO instrument, but
was detectable in the PIN. For the analysis we used the model ``D''
background \citep{fukazawa09a}.  As the PIN background rate is strongly variable around
the orbit, we first selected source data to discard events within 
500s of an SAA passage, we also rejected events with day/night elevation
angles $> 5^\circ$.  The time filter resulting from the screening
was then applied to the background events model file to give PIN
model-background data for the same time intervals covered by the
on-source data. As the background events file was generated using ten
times the actual background count rate, an adjustment to the 
background spectrum was applied to account for this factor.  
{\sc hxddtcor v2007 May} was run to apply the deadtime correction to the
source spectrum. To take into account the cosmic X-ray background
\citep{boldt87,gruber99} {\sc xspec} v 11.3.2ag was used to generate a
spectrum from a CXB model \citep{gruber99} normalized to the $34^\prime
\times 34^\prime$ {\it Suzaku} PIN field of view, and combined with the PIN
instrument background file to create a total background file.  The mean 
exposure times for XIS are given in Table~1. The exposure times for the PIN were  
112, 204 and 59\,ks for 2005 Nov 10, 2008 Nov 6 and 2008 Nov 23, respectively. 

Spectral fits used data from {\sc XIS} 0, 2 and 3, for the 2005 data.
As use of XIS2 was discontinued after a charge leak was discovered
in Nov 2006, the 2008 analysis used only XIS 0 and 3.  In this paper, 
XIS data were
fit over $3.0-10$\,keV. PIN data were fit 
simultaneously with the XIS, in the range 15-50\,keV. 
In the spectral analysis, the {\sc pin} model flux was increased by a factor
1.16 for 2008 data and 1.18 for 2005 data, which are the appropriate adjustments for the instrument
cross-calibration at those epochs of the observation.
XIS data were binned at the HWHM instrumental resolution while PIN data were binned
to be a minimum of 5$\sigma$ above the background level for the  
spectral fitting. 

\section{Spectral Fitting} 
\label{sec:spectralfitting}
The mean \suzaku  count rate over 0.5-10\,keV during the low state of 2005 was
0.45 XIS count\,s$^{-1}$ per  FI XIS and 0.04 PIN count\,s$^{-1}$. During the
high-state of 2008 Nov 6 it was 2.06 XIS count\,s$^{-1}$ per FI XIS 
and 0.07 PIN count\,s$^{-1}$ (Table~1). The count rates recorded correspond to an observed 
2-10\,keV flux range $0.87- 2.42 \times 10^{-11} {\rm erg\, cm^{-2} s^{-1}}$ 
and a 10-50\,keV flux range $2.44 - 3.63 \times 10^{-11} {\rm erg\, cm^{-2} s^{-1}}$.  
The 2005 data represent a historically low state, as noted by 
\citet{terashima09a} and   during 2008 the source was close to the 
historical average flux level.  The 2008 Nov 6 {\it Suzaku} observation was 
accompanied by a contemporaneous HETG exposure that reveals a wealth of emission and 
absorption features from several zones of ionized gas. The HETG 
analysis will be described in detail in another paper \citep{lobban10a}. 

 \subsection{Fitting the PCA Offset Component}
\begin{figure}
\includegraphics[scale=.48,angle=0]{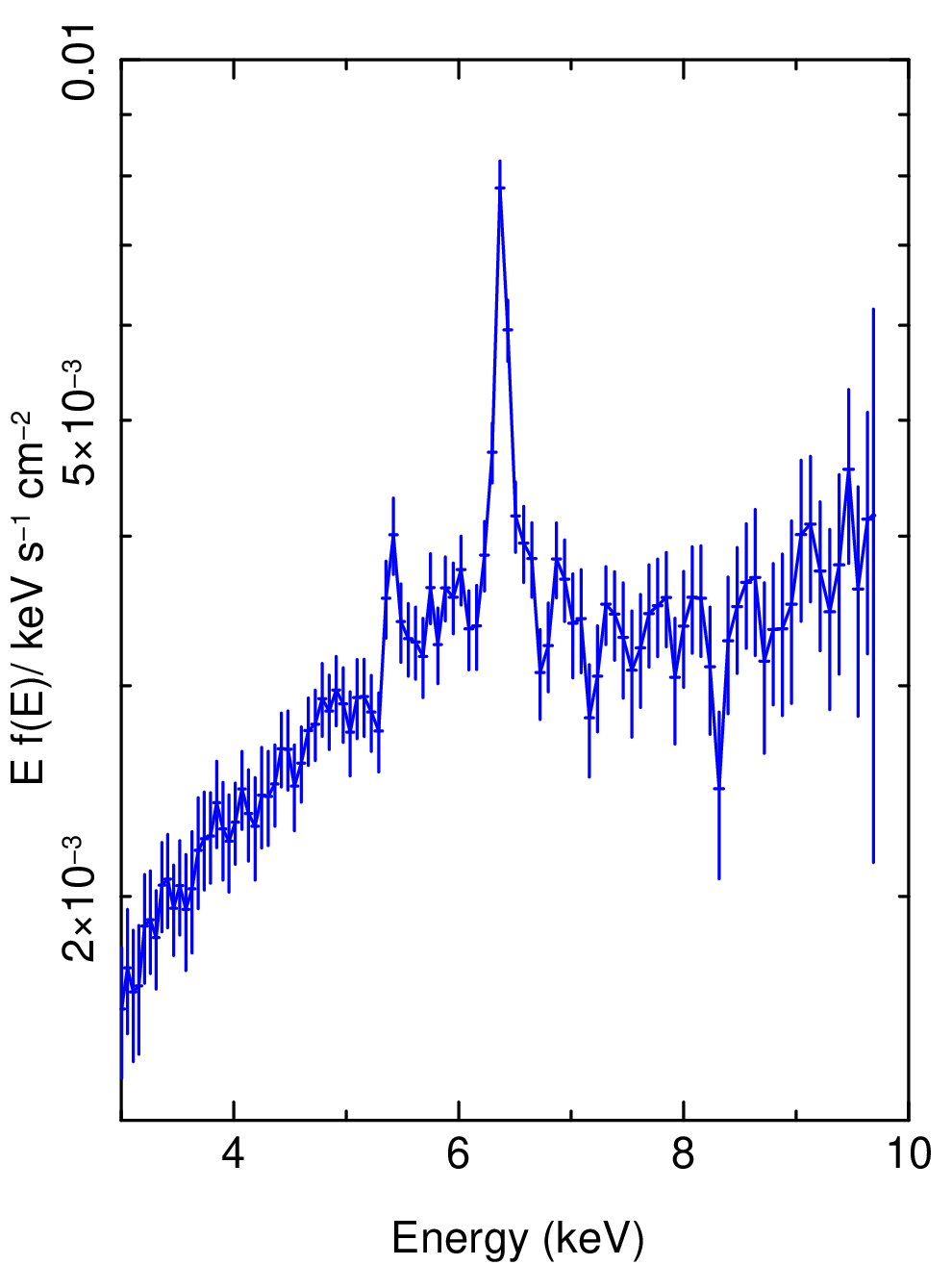}
\caption{The offset component from the PCA decomposition of all of the 2005 and 2008 \suzaku XIS$0+3$ 
and PIN data. A strong narrow Fe\,K$\alpha$ line is evident at 6.4\,keV, 
along with emission at 5.4 and, more weakly,  at 6\,keV. 
 \label{fig:pca} }
\end{figure}

To assist in modeling the full-band {\it Suzaku} data, we first
performed a decomposition of the events from all three {\it Suzaku}
observations using Principal Components Analysis using the SVD method
and code of \citet{miller07a}.  A full description of the PCA
decomposition of these data is reported by \citet{miller09b}.  In summary, PCA is a mathematical decomposition 
of the data into orthogonal 'eigenvectors'. The dominant variations 
are ascribed to eigenvectors of the lowest orders. 
In the case of NGC~4051 the data were found to be 
well-described by  a steady ``offset'' 
component having a hard spectrum (Figure ~\ref{fig:pca} and see \citealt{miller09b}) while the 
first order variable component, eigenvector 1, is 
consistent with an absorbed power-law of 
constant slope whose variations in intensity dominate the spectral
variability of the source (consistent with the analysis of
\citealt{terashima09a}).  The physical origin of the offset component is
of great interest. The low-state spectrum of NGC 4051 is composed of
this component plus some contribution by eigenvector 1.  Fitting the
lowest flux subset of the 2005 data, \citealt{terashima09a} found the
hard component to be best described using a partially-covered
reflection component. Our analysis of the PCA offset component
provides another way to probe the lowest flux levels of the source
and, while we find an acceptable fit using a combination of reflection
(modeled using {\sc pexrav}) plus a contribution from the powerlaw
continuum ($\Gamma=2.3$, fixed from fitting eigenvector\,1), both
emission components require absorption by a complex of ionized
gas. Fortunately the multiple layers of ionized gas can be well
constrained using HETG and LETG data \citep{collinge01a,steenbrugge09a,lobban10a}, 
reducing the 
degeneracy of the {\it Suzaku} fit.  In addition to these components,
Fe\,K$\alpha$ line emission is evident in the offset spectrum at a
fitted energy $E=6.398\pm0.023$\,keV having a width
$\sigma=0.045^{+0.037}_{-0.045}$\,keV and flux $n=1.90 \pm 0.26 \times
10^{-5}$ photons cm$^{-2}\, {\rm s}^{-1}$.   (Throughout this paper,
errors are quoted at 90\% confidence for the appropriate number of
interesting parameters in the fit.) The resulting fit statistic was 
$\chi^2=130/153\, d.o.f$. 
 
Intriguingly, the normalization of the reflection continuum 
is extremely high relative to that of the primary continuum. The 
inconsistency of reflection and continuum strengths may
indicate that the partial-covering absorption comprises a larger
part of this component than currently modeled.  The relatively high
flux of the hard component was also noted by \citet{terashima09a}
based upon fitting the 2005 data, and is fully explored by \citet{lobban10a}.

In addition to the prominent  narrow Fe\,K line, a strong excess of 
line emission is evident in the PCA offset spectrum, at 5.44 keV  
(Figure~\ref{fig:pca}). No significant residuals 
appear in eigenvector 1 at that energy \citep{miller09b}. The new line is thus consistent 
with having an origin  
associated  with the hard offset component and neutral component of Fe\,K$\alpha$ emission.

\subsection{Fitting the 2008 Contemporaneous Chandra and Suzaku data}
\label{sec:hetg_suz}
\begin{figure}
\epsscale{1.0}
\plotone{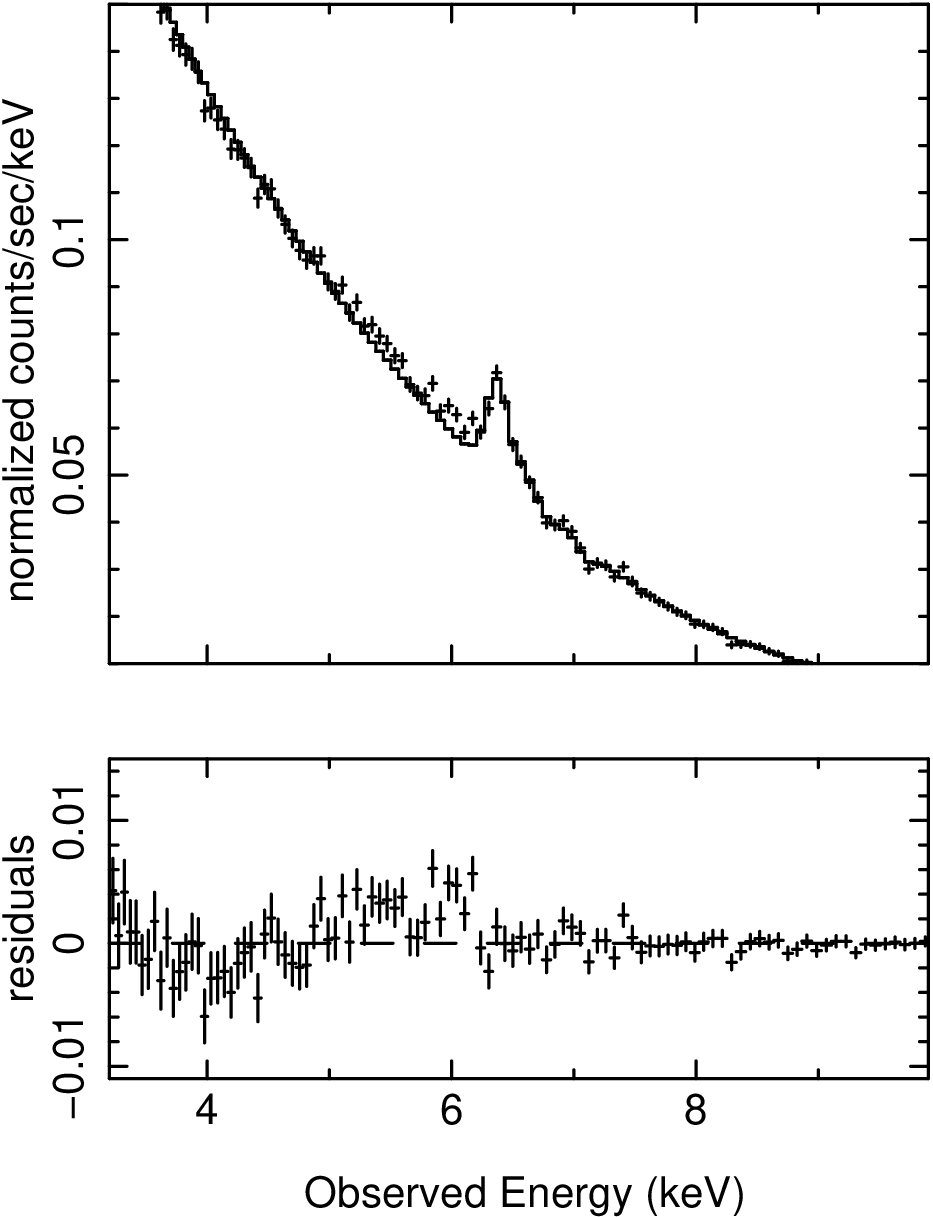}
\caption{\suzaku XIS$0+3$ data and residuals from the mean 2008 Nov 6  spectrum, 
compared to an absorbed powerlaw plus Gaussian line at 6.4\,keV  
\label{fig:cts_res2008}}
\end{figure}

As the 2008 Nov 6 epoch provides contemporaneous HEG data for the {\it Suzaku} observation, we 
fit those spectra  together over 3-8\,keV, combining the superior energy-resolution of HEG 
and the good statistical quality of the FI XIS units (0,3) to obtain the best possible 
constraints  on the width and energy of the neutral component of 
Fe\,K$\alpha$. A joint fit to those data using a simple powerlaw plus single Gaussian line 
 showed  residual excesses  in the 5 - 6 keV band in {\it Suzaku} XIS data 
(Figure~\ref{fig:cts_res2008}). 

We refit, adding to the model a single component of absorption, a Gaussian representation of the  
 K$\beta$ line fixed at a rest-energy of 7.05\,keV with a flux linked  to be 
13.5\% of the K$\alpha$ component \citep[e.g.][]{leahy93a,palmeri03a}, 
a narrow absorption line detected 
at an observed  energy of 7.1\,keV (as discovered by 
\citealt{pounds04c} and confirmed by \citealt{terashima09a})  plus a component to model 
the narrow line emission evident at 6.62\,keV in HEG data \citep{lobban10a}.
The fit yielded 
a line energy 
$E=6.410 \pm 0.015$\,keV, width $\sigma=50^{+26}_{-33}$ eV and normalization 
$n=1.90\pm 0.395 \times 10^{-5} {\rm photons\, cm^{-2}\, s^{-1}}$. 
The equivalent width for the Fe\,K$\alpha$ 
line component  is  95\,eV against the total continuum level observed 
during 2008. 
The constraint on
the width of the Fe K$\alpha$ line is  equivalent to a FWHM velocity-broadening of
$5540_{-3664}^{+2846}$\,km\,s$^{-1}$ (90\,percent confidence interval). 
Assuming  the true line centroid energy to be 6.40\,keV, the 
90\% confidence constraint 
on energy from the combined HEG/XIS data 
limits the bulk velocity of this emitter to  
$250 \ga v \ga -1200$\,km\,s$^{-1}$ where negative velocity denotes outflow. 

As the PCA decomposition indicates the presence of a weak broad component of 
Fe K$\alpha$ on eigenvector 1 \citep{miller09b}, 
we tried adding a second component of Fe K$\alpha$ emission to the model, 
with the energies of both line components linked. This addition improved the fit-statistic by 
$\Delta \chi^2=5$, giving 
 the same line 
energy as for the single component model, with  $\sigma_1 =1^{+23}_{-1}$ eV and normalization 
$n_1=1.12\pm 0.32 \times 10^{-5} {\rm photons\, cm^{-2}\, s^{-1}}$ (EW$=41$ eV),
$\sigma_2 =145^{+70}_{-52}$ eV and normalization 
$n_2=1.07\pm 0.45 \times 10^{-5} {\rm photons\, cm^{-2}\, s^{-1}}$ (EW$=43$ eV).  
Although the two-component parameterization of Fe\,K$\alpha$ 
 offers only a marginal improvement to the fit, it is of interest with regard to consistency with the 
PCA decomposition, which indicates the weak broad component of Fe K$\alpha$
 emission to be present on eigenvector 1. 

In subsequent fits we fix the Fe K$\alpha$ model line width at 
$\sigma=50$ eV obtained from the simple fit, 
 while bearing in mind the possible more complex solution for the line profile.  
As the PCA is consistent with a common origin for  Fe K$\alpha$ emission and 
the line at 5.44 keV, and as the widths of the two lines are consistent (and as there is no 
other useful constraint available for the indeterminate line width) hereafter we fixed 
both to $\sigma=50$ eV  in spectral fitting.

\subsection{Fitting the 2005 Suzaku data} 
\label{sec:linesig}

\begin{figure}
\epsscale{1.0}
\plotone{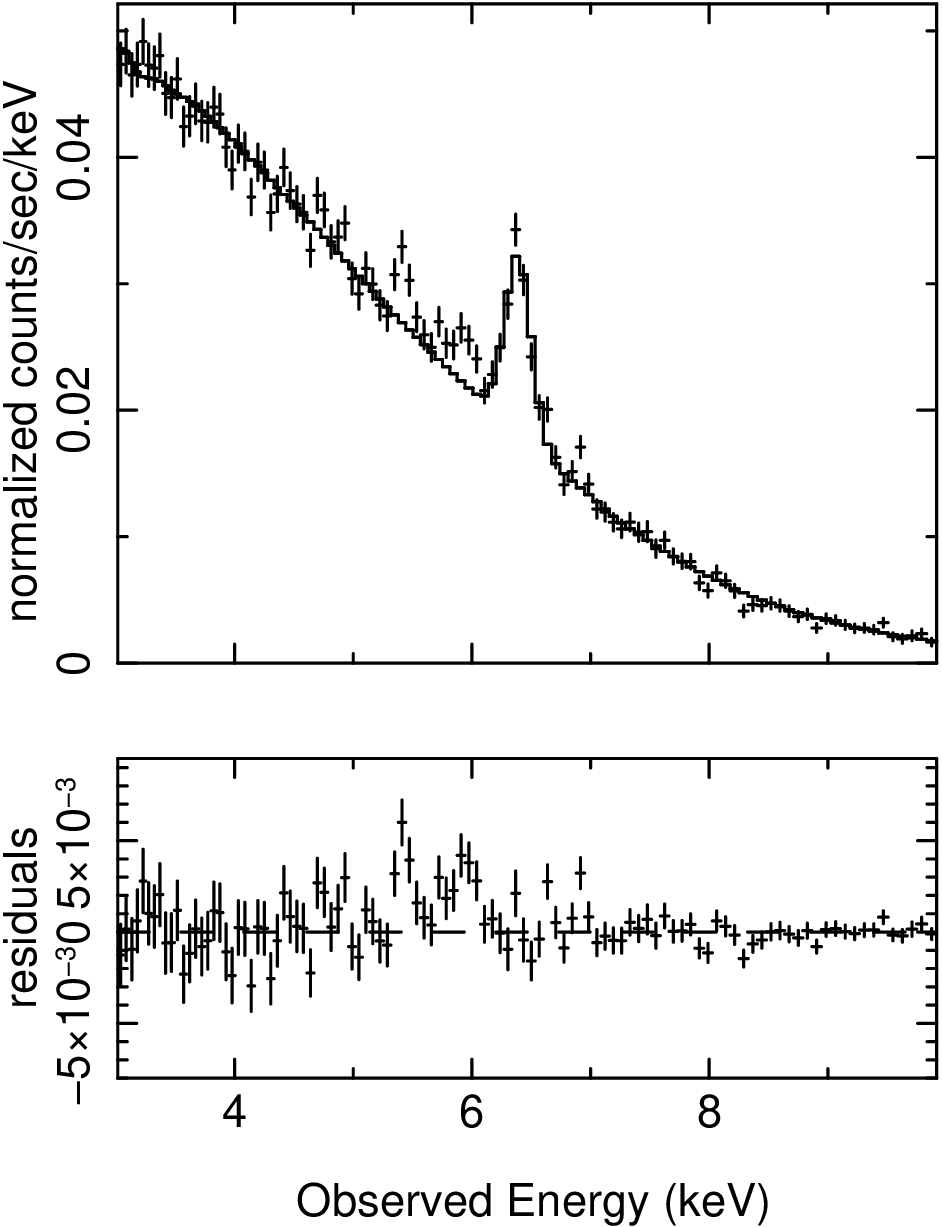}
\caption{\suzaku XIS$0+2+3$ data and residuals from the mean 2005 spectrum, 
compared to an absorbed powerlaw plus Gaussian line at 6.4\,keV, showing the residual excess counts 
at $\sim 5.44$ and $5.95$\,keV \label{fig:cts_res}}
\end{figure}

To assess the significance of the line at 5.44 keV we returned to direct
fitting of the 2005 data, where the source was observed to be at a low
flux and where PCA indicates that the line would have the highest
equivalent width.  We fit the summed 2005 data from the FI XIS units 
0, 2 and 3 using an absorbed continuum model, applying it to the 
3-10\,keV band for the purpose of examining the data for  features
of interest.  A Gaussian line was included in the fit, to account for
the Fe K$\alpha$ emission, fixed at the energy and width found using
HEG/XIS. As evident in Figure ~\ref{fig:cts_res} the mean spectrum
from 2005 data shows an excess of counts at 5.44\,keV with additional structure 
evident around 6 \,keV. We performed a more complex 
fit to properly assess the line strengths and significance. 
We added to the model a Gaussian representation of the  
Fe K$\beta$ line fixed at a rest-energy of 7.05\,keV with a flux linked  to be 
13.5\% of the K$\alpha$ component. The model  also included  
the narrow absorption line detected 
at an observed  energy of 7.1\,keV as discovered by 
\citet{pounds04c} and confirmed by \citet{terashima09a},  plus a component to model 
the narrow line emission evident at 6.62\,keV \citep{lobban10a}.
 The fit yielded $\chi^2=176/101\, dof$ and   
the strongest unmodeled feature remains 
evident as an excess of counts at 5.44\, keV.  

Addition of  a Gaussian line ($\sigma=50$\, eV) reduced the fit-statistic 
by $\Delta \chi^2= 32$, yielding  
$E=5.44 \pm0.03$\,keV,  flux  $n=5.03^{+2.02}_{-2.01} \times 10^{-6}$ 
photons cm$^{-2}{\rm s^{-1}}$ and equivalent width 
$46 \pm 16$ eV. Addition of a second line of the same width 
 yielded a further improvement 
$\Delta \chi^2= 34$, $E=5.95 \pm 0.05$\,keV,  line flux 
$n=5.05^{+2.05}_{-1.95} \times 10^{-6}$ photons cm$^{-2}{\rm s^{-1}}$, 
with an equivalent width  of $44^{+17}_{-16} $ eV and a final fit statistic 
$\chi^2=117/97\, d.o.f.$  
In this fit the flux and equivalent width of the Fe K$\alpha$ line were 
$n=1.59 \pm 0.23 \times 10^{-5}$  and $195 \pm 24$ eV respectively. 

\begin{figure}
\epsscale{1.0}
\plotone{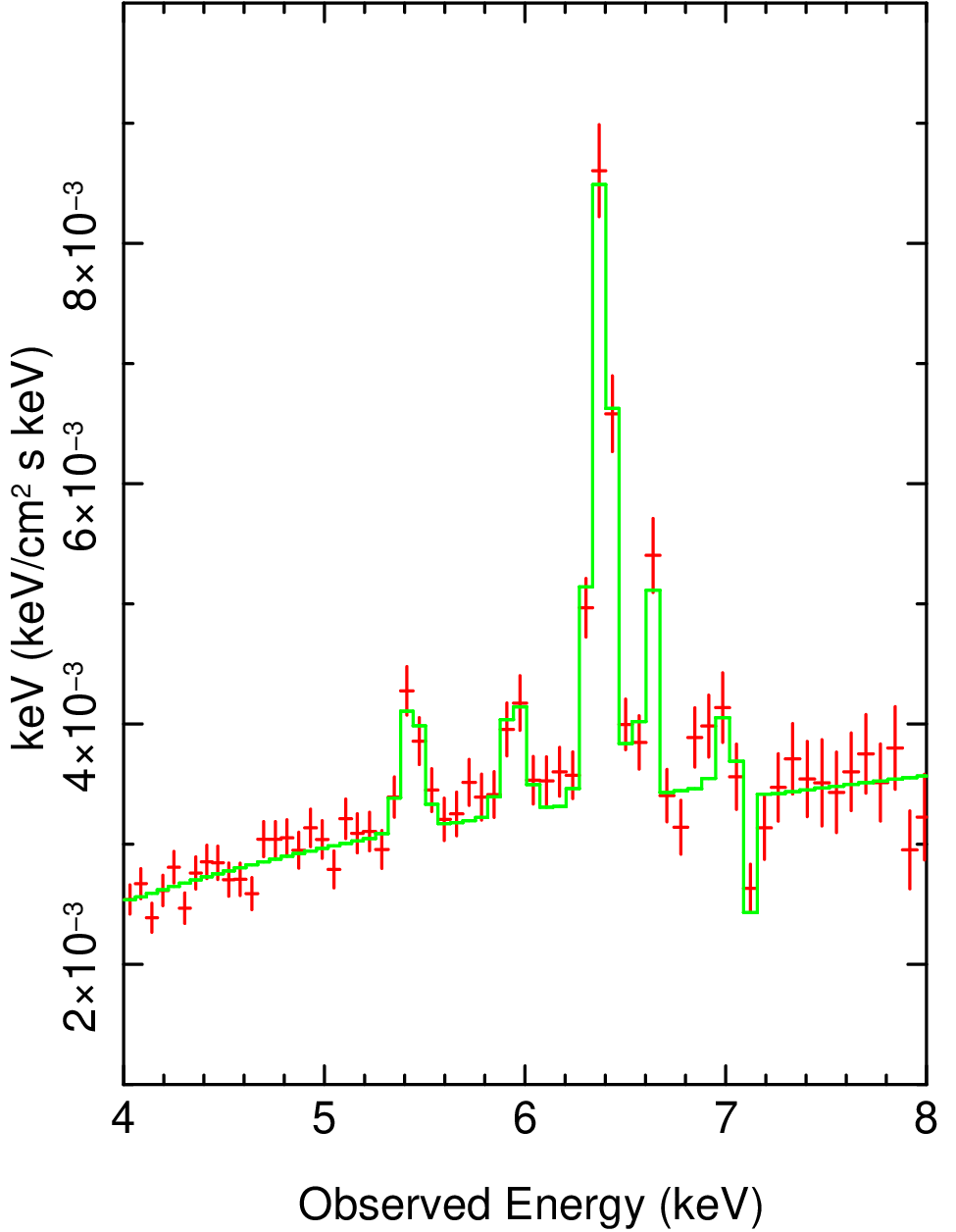}
\caption{The very low state data (red crosses) from within the 2005 observation 
shown with the model line (solid green), illustrating the emission 
and absorption lines present in the data. 
 \label{fig:vloeeuf}}
\end{figure}

To further examine the source behavior at low flux levels we 
isolated the lowest  flux subset  of the 2005 data  using an 
intensity filter based upon the 0.5-10 keV band count rate,  
taking data below a threshold of 0.47 ct s$^{-1}$ (per FI XIS).  
 The intensity cut   effectively removed the periods of 
source  flaring from consideration. This filtering thus resulted in a 
very-low-state spectrum that had a mean 2--10\,keV flux 
$6.1 \times 10^{-12} {\rm erg\, cm^{-2} s^{-1}}$. Again, fitting a simple 
absorbed continuum model over 3--10\,keV leaves two strong
positive residual features between 5--6\,keV (Figure ~\ref{fig:vloeeuf}) 
confirming the prominence of the 5.44\, keV line at low flux levels and indicating the possible presence of 
emission at 5.95\, keV.  

\subsection{The Full Model}

\begin{figure}
\epsscale{1.0}
\plotone{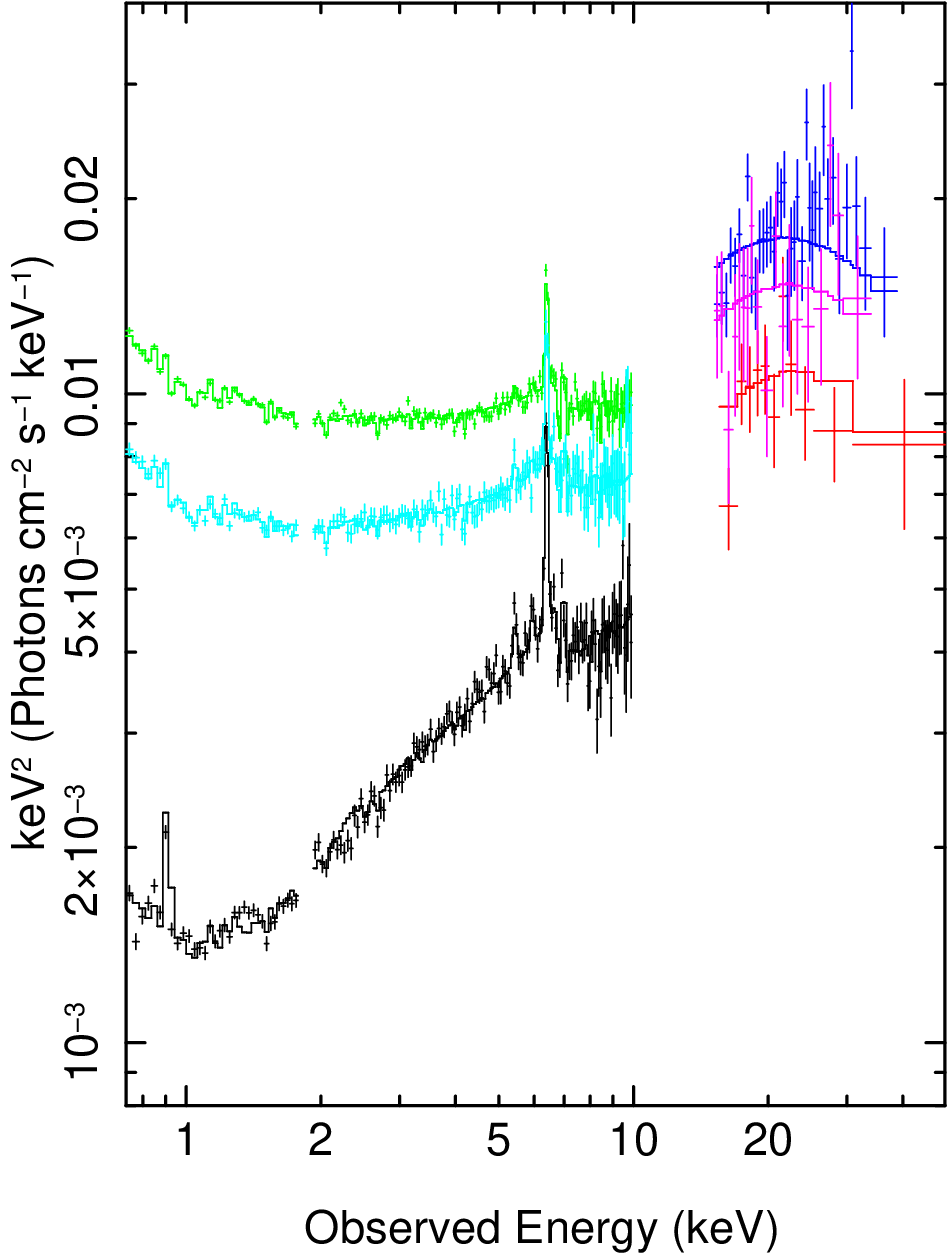}
\caption{The 2005 (XIS is black, PIN is red), 2008 Nov 6 (XIS green, PIN ) 
and Nov 23 (XIS pale blue, PIN magenta) 
data along with the broad-band model described in the text. 
 \label{fullspec}}
\end{figure}

To assess the full properties of the source and investigate the robustness of the line detections 
to the continuum model used, we then fit over 0.75-50 keV (the full band of the {\it Suzaku} data). 
We  simultaneously fit 
spectra from  the three {\it Suzaku} observations.  The 2008 
HETG exposure (that overlapped the 2008 Nov 6 {\it Suzaku} 
observation) shows a wealth of absorption and emission lines that can
be modeled using three zones of ionized gas \citep{lobban10a}:  those 
soft-band absorbers 
were fixed in the {\it Suzaku} fit, after which two
additional gas layers were required to accurately model the source
across the full {\it Suzaku} bandpass.  The fit also included an Fe K$\alpha$ emission
line fixed at a width of $\sigma=50$\, eV as before. 
Finally, an absorbed ionized
reflector was included to account for the curvature in the soft band.
The full model is detailed in Table~2 and the fit is illustrated in
Figure~\ref{fullspec}.   The marked spectral 
variability observed (Figure~\ref{fullspec}) can be accounted for by
allowing variations in the covering fraction of one absorbing layer
(having $N_H \sim 10^{23} {\rm cm^{-2}}, {\rm log}\, \xi \sim 0.18$),
a solution that is similar to other well-studied AGN of this class
\citep[e.g.][]{pounds04c,risaliti07a,miller08a,turner08a}.  This fit
yielded $\chi^2=947/547\, d.o.f.$ with the overall curvature modeled
well across all epochs and the residual $\chi^2$ contributed mainly by 
some unmodeled spectral features.  The addition of a line ($\sigma=50$\, eV)
 constrained to lie in the range  3 - 8 keV yielded a detection at 5.44$\pm0.03$  keV  
 and improved the fit statistic by $\Delta \chi^2=20$; addition of another 
line  under the same constraints gave 5.98$\pm0.05$ keV and 
$\Delta \chi^2=9$ improvement to the fit.   
We conclude that the high significance for the detection of the 5.44\, keV line is robust to the
continuum used while the significance of the detection of a line at 5.95 keV is sensitive to
the continuum form assumed.

\subsection{A Critical Examination of the Reality of the Lines}

Before proceeding  any further with the  modeling,  
we performed several checks to be assured that the new line is  
attributable to the AGN and is significant. 
First, to confirm that the line is not an artifact of poor background subtraction, 
we examined the background spectrum and calibration source data. 
First we note that the background comprises just 2.5\% of the total count 
rate in the 5-7 keV band  for the XIS 
spectral data when the source is at its lowest flux level, during 2005 and 
1.3\% when the source is brighter, during 2008.  
Regarding the line at 5.44\,keV, we found there to be no 
line emission evident at that energy in the background spectrum. Fitting the background spectra from 
the three {\it Suzaku} observations we obtained an upper limit (90\% confidence) for the 
flux of a line at 5.44\, keV in the background spectrum 
$n < 4.14 \times 10^{-8}$ photons cm$^{-2}{\rm s^{-1}}$, i.e. $< 1$\% of the detected line flux. 

\begin{figure}
\epsscale{1.0}
\plotone{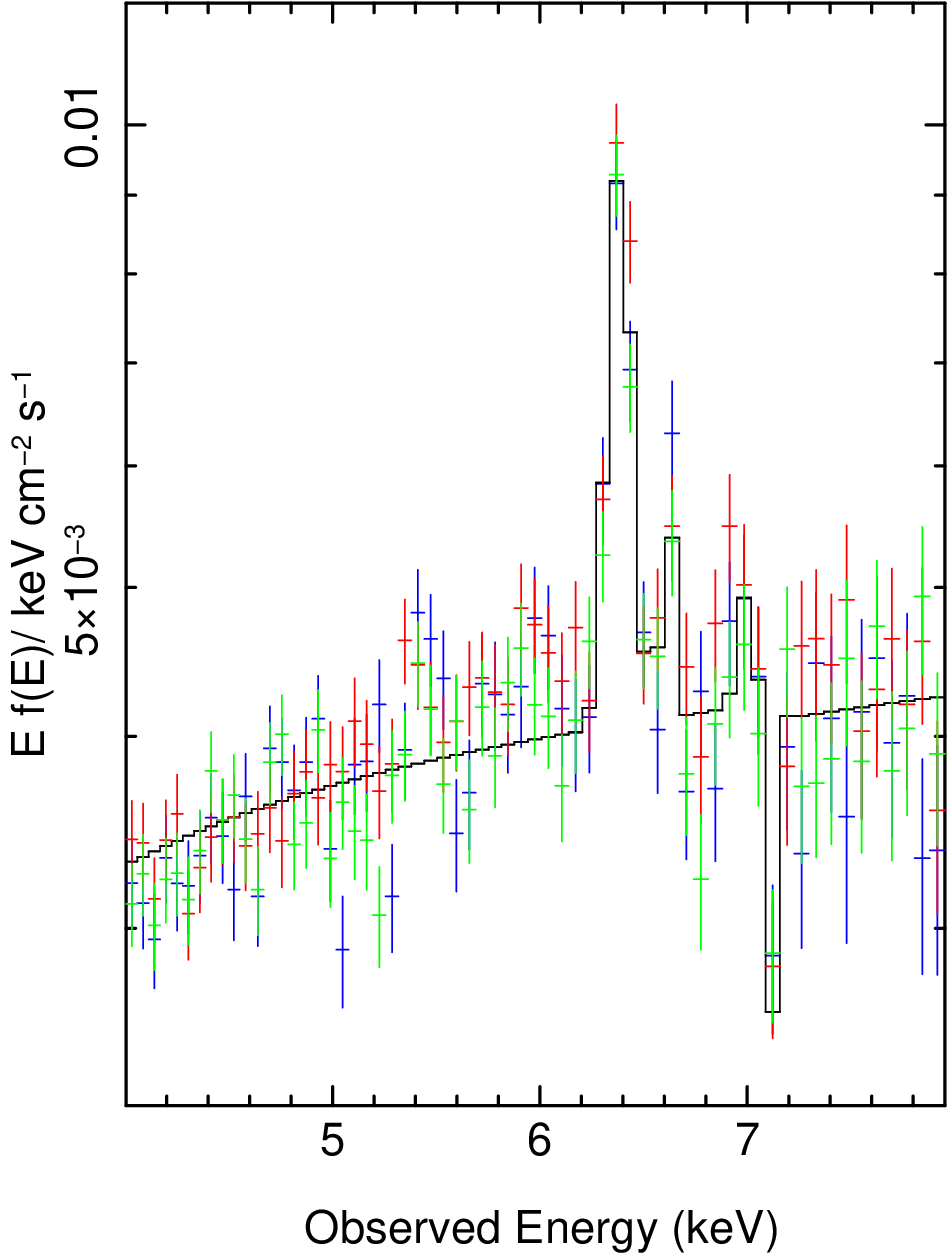}
\caption{The 2005 data for the three XIS detectors against the continuum 
model (black line); 
blue is XIS0, red is XIS2 and 
green is XIS3 
 \label{fig:in3xis}}
\end{figure}

The lack of a feature at comparable 
flux or equivalent width in the background data  also rules out an origin of the 5.44\,keV line 
as a detector feature. We then examined each XIS independently, and found the 5.44\,keV  
line to be significantly detected in each XIS unit independently, further verifying the reality of  
the reported feature (Figure ~\ref{fig:in3xis}).   

Regarding the more tentative line at 5.95\,keV: the XIS chip does include a Mn calibration source 
emitting a line at 5.9\,keV for the purpose of calibration of the XIS energy scale. The calibration  
source is located at the chip edge and the counts from that source are obviously 
conservatively excluded from any source and background 
extraction cells for scientific analysis. However, we  do detect a  weak 
 Mn K$\alpha$ emission in the background spectrum. The Mn contamination is  
at a level $< 10\%$ of the measured line strength in the AGN spectrum. As the source 
extraction cell is further from the Mn calibration source than the background cell and 
therefore subject to less contamination,  
we estimate the Mn line contamination from the calibration source to be $< 10$\%  for the 2005 spectrum. 
In spectral fitting, the greatest concern for the Mn line is the possible contamination by the weak broadened Fe K$\alpha$ 
emission evident in the fit to eigenvector 1 \citep{miller09b} which may lead to the strength of the Mn 
line being over-estimated from simple fits. Because of the various 
issues associated with a clean measurement of any line 
at 5.95\, keV and the sensitivity to the continuum form assumed 
we concentrate only on the strong line detection at 5.44\, keV. 

To confirm the significance of the line  
we performed Monte Carlo simulations using the method described in 
\citet{porquet04a} and in \citet{markowitz06a}. We took the 
 null hypothesis to be that the 
 spectrum is simply an absorbed power-law continuum with 
 parameters derived from fitting the broad-band data but allowing for the 
statistical uncertainty on the continuum parameters and including the 
narrow Fe K$\alpha$ line 
whose presence is well-established in this source 
\citep{lobban10a}.  
 We used the {\sc xspec} command {\it fakeit} to create 3000 fake 
{\it Suzaku} spectra  with photon statistics expected
from the 2005 exposure, assuming the same  instruments to be  operational as 
for the actual observation. The simulated data were grouped to the 
HWHM energy-resolution of the instruments, the same as the observational data. 
 Following the procedure used to test the real data for
the presence of a narrow line, we fitted each fake spectrum  to obtain the 
 values  of $\Delta \chi^2$ obtained from statistical fluctuations 
in the data. 
To map the distribution of $\Delta \chi^2$  across 
the simulated spectra, each simulated spectrum was fitted over the
3-10 keV energy range, stepping through using  energy bins whose centers were increased 
in increments   
of 100 eV.  The line energy was allowed to be free within each energy bin tested 
 and the value of  $\Delta \chi^2$ was recorded at each point in 
the spectrum. This method  makes no assumptions about the energy at which a line 
might be detected and over the course of the testing, all energies are tested for the presence 
of a line. 
When we fit the simulated data we are  testing whether we 
can produce, from statistical fluctuations, a contribution to $\chi^2$ at the same or greater level 
as found in the actual data  at any energy in the range of interest. 
The fits to the simulated data  yield a  
distribution of $\Delta \chi^2$  for 
comparison with the actual data. 
We found  the most extreme statistical fluctuation to 
yield a $\Delta \chi^2$ contribution of 19.3 in the 3-10\,keV band 
in the set of simulated data (i.e. in 3000 simulations no false line appears at any energy contributing 
$\Delta \chi^2=32$ as found in 2005 data). 
Thus the probability of satisfying the null hypothesis is 
$p < 3.3 \times 10^{-4}$ for the line found at 5.44\, keV.

\subsection{Application of a Disk Hotspot Model}

As an alternative to modeling  using individual Gaussian
lines, we fitted the data using a single disk line from a narrow annulus, such that the red horn of such a line 
might explain the peak at 5.44\, keV. We assumed 
the system to contain a non-rotating black hole, that the line is Fe K$\alpha$ emission from neutral
material (at 6.4\,keV), that the emissivity pattern across the disk can be described
by $r^{-q}$ where the emissivity index q=-2.5 and that the hotspot
exists over a narrow annulus of width $\Delta r=1r_g$.  We  
used the same baseline model as for testing the Gaussian lines 
(section~\ref{sec:linesig}).  

To fit the 5.44\, keV peak  as the red  Doppler horn of a disk-line  
requires an emitting radius  21$\pm 3 r_g$  
(where $r_g = GM_{BH}/c^2$ denotes gravitational radii) 
in  a low inclination system 
with  ${27^{+1}_{-3}}^\circ$ with 
$\chi^2=128/95\, d.o.f$.; the corresponding blue horn is 
then predicted to lie at 6.62\,keV 
and the data are consistent with that model. 
While there is evidence for line emission  in the 6.5-7\,keV regime, such lines 
 are commonly observed as emission from ionized species of Fe, 
and thus the identification of 
emission blue-ward of 6.4\,keV  is currently ambiguous.

\subsection{Examination of the Line Variability}
\label{sec:linevar}   

\begin{figure}
\epsscale{1.0}
\plotone{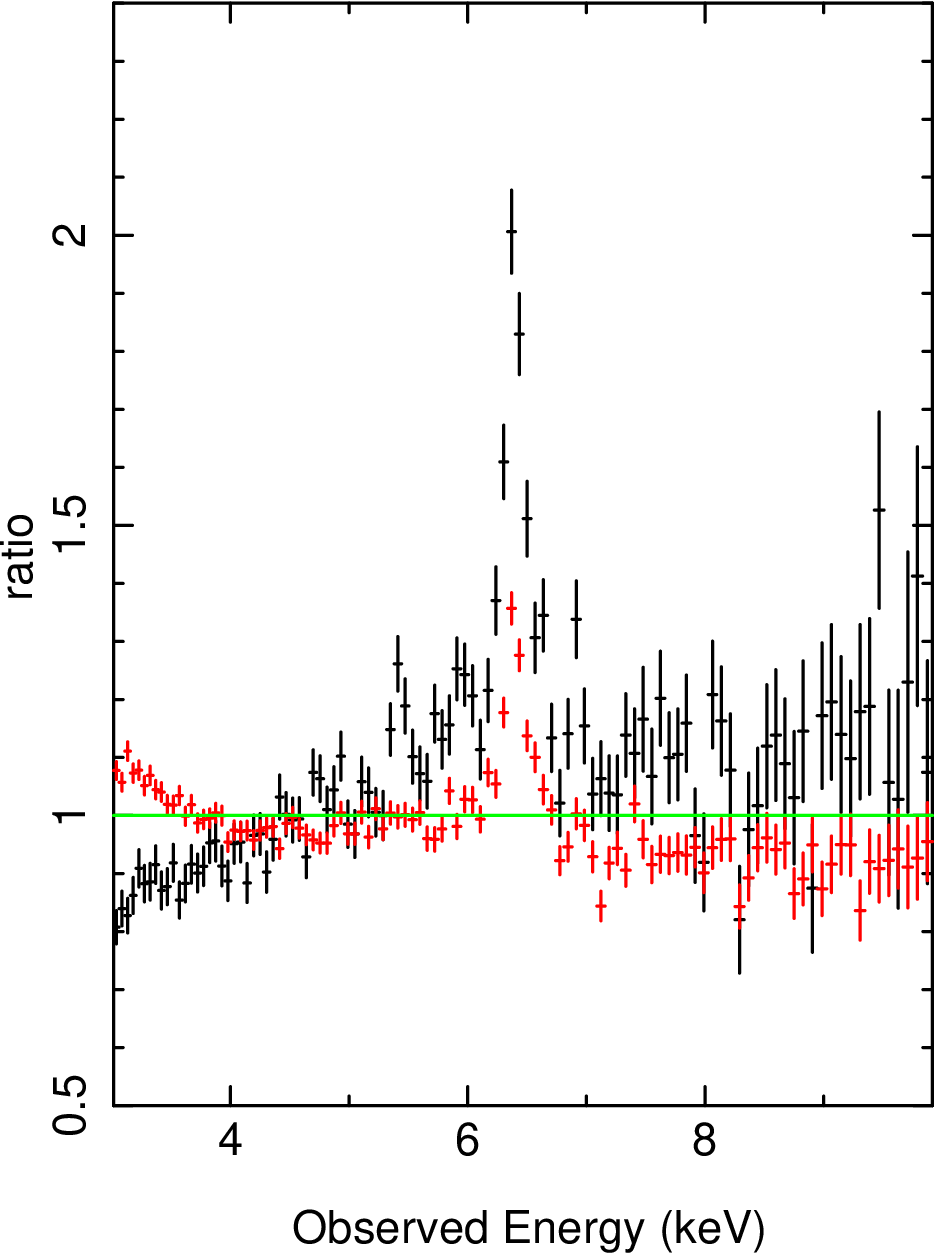}
\caption{{\it Suzaku} data shown as a ratio to the mean local continuum model.  
The red line is the  data from 2008 Nov 6 and the black line is the 
 data from 2005. 
 \label{fig:hiloratio} }
\end{figure}

Fits to the mean spectra from each observation have shown consistent 
line fluxes for Fe K$\alpha$ and the line at 5.44\, keV  
and these show an  equivalent width that  appears to have 
changed as the source flux varied.  
Figure~\ref{fig:hiloratio} shows the ratio of data in the Fe K$\alpha$ regime to a common local 
continuum fit; the fact that the source spectrum is steeper at high flux 
is reflected in the systematics of the residuals.  To confirm the 
significance of the change in equivalent width we fit the 2008 data with a model that fixed the 
line equivalent width for the feature at 5.44\,keV to that  found during 2005; 
after refitting this resulted in a worse fit with $\Delta \chi^2=142$. 
Alternatively, fixing the flux of the new lines at the 2005 values and refitting, 
we found $\Delta \chi^2=0$, indicating that the line fluxes may be consistent 
with lines of constant flux across the data. 
 
As the limits on line flux provide the potential to
distinguish between models, we tested all available high-quality data
in a self-consistent manner.  Taking the model from
section~\ref{sec:linesig} we fixed the line energy and width to
specifically test the constancy of the 5.44\, keV line. 
In addition to testing the mean spectrum from  each {\it Suzaku} 
observation, we sub-divided the 2005 and 2008 exposures to sample the 
line more finely in flux.   
For 2005 an intensity selection was made on the 0.5-10 keV count rate 
at 0.47 ct s$^{-1}$ per XIS as before.  For 2008 intensity 
selections were made $ < 2.0 $ ct s$^{-1}$/XIS ({\it i1}), 2.0 - 3.0 ct s$^{-1}$/XIS ({\it i2}) and 
$ > 3.0$ ct s$^{-1}$/XIS ({\it i3}). The results are shown in Table~3 
where both the mean fits for each observation are tabulated, as well as the 
intensity-selected results.   As noted previously, the line is required 
at a high level of confidence in 
the 2008 Nov 6 data (which is more sensitive to the features than the 
Nov 23 data, owing to the long exposure and high flux state of the source). 

Table~3 shows line flux along with improvements in the fit-statistic and
equivalent width for each fit.  Note that the mean observation 
fits in Table~3 show 
slightly tighter constraints in line flux for 2005 compared to the values noted
in section~\ref{sec:linesig} because  the initial fits to 2005 data had the 
line energy left free.  
The data are consistent with the 5.44\, keV line existing
at the same flux level throughout the {\it Suzaku} observations considered in Table~3 and the high significance 
of the line detection across several time slices of data provides compelling evidence for the reality of 
the line; conclusively 
ruling out the possibility of the line detection  being a statistical fluctuation in the spectral data. 
Considering the three independent  {\it Suzaku} detections of the line, as shown in Table~3 {\it Suzaku} 
observations 1,2 and 3 yield improvements to the fit $\Delta \chi^2=32,19,19$ respectively and 
a probability of all three detections being false is $p < 3 \times 10^{-11}$.   

We repeated the fits with the line energy allowed to be free and found
the fitted energy to be consistent with 5.44\,keV and 
that the line flux had not been significantly affected by freezing
the energy.  

To extend the test for line flux variations we reduced and fit the
{\it XMM-Newton} spectra obtained during 2001 and 2002. We 
followed  the standard reduction method for the pn data,  as
detailed by \citet{ponti06a} and found the data to be consistent with the presence of a line at the same 
flux as found using {\it Suzaku}.  
An independent analysis of the {\it XMM} data by \citet{demarco09a} 
reported the presence of an excess of counts in the 5.4-6.2 keV band for NGC 4051, compared to their 
parameterization of the local continuum. Examination of the {\it XMM} data shows an excess of 
emission at $\sim 6$\, keV in the {\it XMM} spectra and so tentatively supports the possibility 
of that additional energy-shifted line  in this source. 

{\it BeppoSax}
also observed NGC 4051, finding the source to be in a very low flux
state during 1998 as reported by \citet{guainazzi98b}.  We extracted
and fit the archived {\it BeppoSax} Medium Energy Concentrator
spectrum from units 2 and 3 (combined) and tested for the presence of
the line. However, the data did not yield a significant detection of
and the upper limits on line flux were very loose $\sim 4 \times 10^{-5} {\rm photons\, cm^{-2} s^{-1}}$ for a
line at 5.44\,keV. 
Given the poor constraint obtained we do not consider those data any further.

Combining the results from {\it Suzaku} and {\it XMM} 
observations (specifically, Table~3 lines 1,2,4,5,7-9 \& 10) 
and comparing the data to a constant model yields 
$\chi^2=4.7/7\, d.o.f.$  Further to this test, we split the data by time 
instead of intensity and repeated the test, reaching the same conclusion, 
i.e. that  the data are consistent with a 
line of constant flux over a timescale of several years and over large 
changes in observed continuum flux. The limits on measured line fluxes mean we 
can rule out line variability greater than a factor $\sim 2$ in the line flux  
sampled on these timescales,  across the baseline time 
period considered. 

\section{Discussion}

While numerous claims exist in the literature for emission lines at
 unexpected energies 
\citep[see][for a review]{turner09a}, 
the reality of the lines has been questioned by some. 
\citet{vaughan08a} considered 38 published results on transient emission
and absorption lines reported in the literature: those authors find a
linear relationship between the fitted feature strength and its 
uncertainty. \citet{vaughan08a} noted that observations with more signal apparently
reveal weak lines but do not show tightly constrained strong lines
which should sometimes also show up by chance.  The conclusion of the
\cite{vaughan08a} literature review was that there is a publication
bias in reporting of these results, and that many of the reported
detections are merely statistical fluctuations. This question has now
been addressed  with two systematic analyses of
samples of AGN.  \citet{tombesi10a} present a study of the occurrence
and reality of energy-shifted absorption lines in a sample of AGN
finding a deviation from the linear relationship of line equivalent
width (EW) and uncertainty found by \citet{vaughan08a} in the sense that 
their distribution showed more significant detections of lines in 
sources studied. 
 \citet{tombesi10a} compare their absorption 
line measurements with those of the Fe K$\alpha$ emission in the same
sources and show that these two sets of measurements 
follow the same distribution in the EW/uncertainty plane.  
\citet{tombesi10a} conclude that the 
absorption line detections are generally not statistical fluctuations 
and that the absence of well-constrained detections of strong features
in the \citet{vaughan08a} analysis may be due in part to a limit on
the ability of current X-ray instruments to detect such lines.  It
also would appear likely that long observations of bright sources may
not have been proposed or approved early in the {\it XMM} and 
{\it  Suzaku} missions and so the sample of observations completed to date
is likely biased against those that would have  shown  
tightly constrained energy-shifted lines: such a bias may explain at least some of 
the effect discussed by \citet{vaughan08a}. Another recent study by
\citet{demarco09a} undertook a systematic analysis of a sample of
bright Seyfert 1 galaxies and confirmed many of the individual
detections of energy-shifted emission lines claimed in the literature,
supporting the general reality of the phenomenon by consideration of
the statistics of the sample results as a whole. With conflicting 
views in the literature it is clear that the detection of significant 
new examples 
of the energy-shifted line phenomenon is very important at this time.

\subsection{Hard X-ray Line Emission in NGC 4051}
\label{sec:hardxrayline}

NGC 4051 can be modeled using a powerlaw continuum covered by multiple zones  of  gas, several in the  
column density range $10^{23} - 10^{24} {\rm cm^{-2}}$, 
that impart emission and absorption features to the X-ray spectrum. The marked spectral variability 
with observed flux can be explained by changes in covering of the powerlaw continuum by one of the high-column absorbers, as found for 
other similar AGN \citep[e.g.][]{pounds04c,risaliti07a,miller08a,turner08a}. 
Significant line emission has been observed at 5.44\,keV in 
the 2005 {\it Suzaku} observation of  NGC 4051. Comparison of 
low and high-state X-ray data for NGC 4051 shows that 
the newly-discovered line appears  prominent 
in the low-flux state along with the  narrow component of Fe K$\alpha$ emission from neutral gas, as expected if the observed 
low state is simply those  times when a relatively 
large fraction of the 
continuum is suppressed by  absorption. 

The most recent measurements indicate NGC 4051 to have a 
black hole mass $ M_{BH} = 1.73^{+0.55}_{-0.52} \times 10^6 M_{\rm \odot}$ 
and a radius for the $H\beta$ broad line region (BLR) 
$R_{BLR}= 1.87^{+0.54}_{ -0.50} $  light days \citep{denney09a}. 
The $H\beta$ FWHM in the rms spectrum of \citeauthor{denney09a} is 
$1034 \pm 41$\,km\,s$^{-1}$ (although those authors used the velocity  
dispersion rather than the FWHM in their mass estimate). 
If we assume the same geometrical correction factor between line width 
and circular velocity as those authors to scale the respective FWHM measurements, 
the  line width of Fe K$\alpha$ is indicative of  an origin at 
$r \simeq 1.87(1034/5540)^2 \simeq 0.065$\,light days, or $2 \times 10^{14}$\,cm, 
with a 90 percent confidence range
$8.6 \times 10^{13}- 1.7 \times 10^{15}$\,cm. As we have scaled to the optical reverberation
results, this radius estimate is secure provided the X-ray and optical line-emitting regions
have similar structure and orientation with respect to the observer, but the large uncertainty
is dominated by the uncertainty in the line width measurement. 
Further to the measurement error is the uncertainty 
as to whether, as suggested by the form of eigenvector 1, the Fe K$\alpha$ emission has both broad and narrow 
components, in which case  we would have to conclude that we are  seeing contributions from 
regions both within and outside of the region noted above. 
Given the limited signal-to-noise in the regime of Fe K$\alpha$  in the HEG data 
this question remains open with current data. 

In the single component model for Fe K$\alpha$ emission, spectral fitting to the HEG plus {\it Suzaku} data 
yields an Fe K$\alpha$ emission line of flux 
$n=1.90 \pm 0.40 \times 10^{-5}$ photons cm$^{-2}$\,s$^{-1}$. 
The strength of the Fe K$\alpha$ line emission, normalized to the illuminating continuum,  
can be used to set limits on the reprocessing gas in which it arises. 
Following \citet{yaqoob10a} and correcting for the continuum slope found here, we
estimate an efficiency for the production of line photons (defined as the ratio of
line flux to incident flux above the ionization edge)
to be 
$x_{\scriptscriptstyle {\rm FeK}\alpha} \sim 0.018$. The line measured here sets a lower 
limit on the column 
density of the emitting region 
$N_H > 10^{24}{\rm cm^{-2}}$ and  in the 
toroidal reprocessor  model suggests a global covering factor  
$\sim 0.9$ with an approximately face-on view down the pole of the structure. 
As it is difficult to determine the intrinsic continuum strength from the observed continuum 
the interpretation of these data in the context of the toroidal model is subject to some 
uncertainty that in turn, leaves the derived global covering factor uncertain by a factor of a few. 
  
Combining the minimum column density of the line-of-sight gas
with the radial constraints and knowledge of the gas having a high covering fraction 
yields a mass estimate $\ga 4 \times 10^{-4}$M$_{\odot}$ for the gas emitting the Fe K$\alpha$ line,
assuming the nominal radius of emission of $2 \times 10^{14}$\,cm.
The 90\,percent confidence uncertainty in the distance translates into a confidence region on the 
mass lower limit of $5 \times 10^{-5} - 0.1 $\,M$_\odot$. 
If a torus is not the true geometry of the gas then, of course, 
the covering factor could be different to this value. 
Further to the uncertainties mentioned, the line may be be comprised
of contributions from two regions. Taking instead the two-component
fit to the Fe K$\alpha$ profile then the column and/or global covering
requirements are reduced for each of the two emitting regions.

The PCA decomposition \citep{miller09b} suggests a link between 
 the origin of the Fe K$\alpha$ line and  that of the 
line at 5.44\,keV.
Observation of similar lines  in other AGN has motivated 
discussion of several possible origins, including spallation and hotspot 
emission from  the accretion disk.  

The solution found in the context of the disk hotspot model 
suggests that the emitting radius
is between $18-24\,r_g$.  The orbital timescales at
these radii are   $\sim 4-6$ ks for
18-24 $r_g$ for the black
hole mass considered here. The observation of a steady line flux over
2005 - 2008 provides a constraint on the disk hotspot hypothesis as
the three year baseline is equivalent to $\sim 16,000-24,000$ orbits about the
black hole over the radial range of interest (and tens of orbits
just considering the line persistence within the 2005 observation).  
Theoretical modeling indicates that hotspot events are not expected to
last longer than a few orbital timescales at these small radii
\citep{karas01a} and that a given hotspot would suffer measurable flux
and energy changes as the material spirals in \citep{dovciak04a}.
However, persistent steady lines could arise from a special radius in
the disk or other rotating reprocessor without any constraining
expectation of flux or energy variability. If the special radius is
interpreted as the truncation radius of the inner disk this 
places  the innermost edge of
the disk at 18-24$r_g$. The inner edge of the disk may emit more strongly
than the rest of the structure if it is inflated due to radiation
pressure as in the advection-dominated accretion flow scenario. 
Assuming the  bolometric luminosity to be  $L_{bol}= 10^{43} {\rm erg\, s^{-1}}$ 
(\citealt{vasudevan09a}, corrected to the Tully-Fisher distance 15.2\,Mpc, \citealt{russell04a}) and
assuming a  radiative efficiency $\eta^{\rm BOL} = 0.05$ the   mass accretion rate is estimated to be 
$\dot{M} \simeq 0.0035$\,M$_\odot$year$^{-1}$, $\sim 10$\% of the Eddington accretion rate. 
For such high accretion rates 
the transition from thin disk to an advection-dominated accretion flow 
 may occur anywhere up to a radius of $\sim 2000 r_g$ \citep{narayan98a}  
and so our fitted radius is consistent with such a scenario.   
 If features 
on the truncated edge of the disk are steady in flux, as found 
for the line in NGC 4051, 
then Galactic binaries would also be expected to show 
such lines in the low flux state and the absence of such would disfavor the 
truncated disk origin for the lines. While similar  
lines have not been reported to date for Galactic 
black hole binaries \citep[see][for a review]{done07b}, 
constraints on such lines have not yet been explored in  that source class,   
leaving this an open question. 

In an alternative model, the fact that the specific energy of the 
line is coincident with K$\alpha$ emission from neutral Cr (5.4\,keV)  
prompts a renewed interest in the 
spallation of Fe as a mechanism for enhancing otherwise weak lines, especially
since PCA is consistent with 
a common origin for the new line and the neutral
component of Fe.  In a companion paper, \citet{turner09c}, explore in
detail the spallation interpretation of the new result, extending
the work of \citet{skibo97} in the light of new understanding about
the environs of active nuclei.  \citet{turner09c} find the observed 
abundance enhancement to be high and that these extreme enhancement effects are
most likely to be achieved in gas out of the plane of the accretion
disk.  In such a picture, the timescale for spallation may be as
short as a few years if the cosmic ray output is comparable to the
bolometric output of the nucleus.  \citet{turner09c} also estimate the
expected radio and $\gamma-ray$ flux from the proposed spallation
process in NGC 4051 and find predictions to be consistent with current
flux measurements in those bands.

\section{Conclusions}

Our analysis of \suzaku data from NGC 4051 taken during 2005 and 2008
has revealed line emission at 5.44\,keV  in the
rest-frame of the galaxy.  We have established the reality of the
line at $>99.9\%$ confidence in data from 2005, supported by Monte Carlo
simulations that show the probability of the line being a statistical
fluctuation is $p < 3.3 \times 10^{-4}$. The possibility of the line arising from 
a statistical fluctuation in the spectral data has been firmly ruled out 
by establishing its detection in time-sliced data. Further to this,  
we have confirmed the
line to be evident in all three XIS units independently,  and established 
that the observed line is inconsistent with arising from the X-ray background. 

The source spectrum varies with flux, and the low state is dominated 
by a hard spectral form with the new line plus the neutral component
of Fe K$\alpha$ emission superimposed upon that, suggestive of a
common origin for both. The line has an equivalent width during
2005 of about 45 eV  while the Fe K$\alpha$ line is measured at 195 eV. These 
reprocessed signatures show up prominently in the source low state when the 
continuum is suppressed by the highest covering fraction of absorption. 

  As disk hotspot emission would 
be expected to vary in flux and energy over relatively short
timescales, the limits on line variability disfavor this particular  origin although
the data remain consistent with emission from a special location such as the
innermost radius of the accretion disk.  The alternative picture, that
the line is Cr {\sc i}  K$\alpha$ emission following
spallation of Fe is also found to be a good explanation
of the data: that possibility and its implications are explored in a 
companion paper.

\acknowledgments

TJT acknowledges NASA grant NNX08AL50G. LM acknowledges STFC grant number PP/E001114/1. 
We are  grateful to the anonymous referee whose comments significantly 
improved this manuscript: we also thank 
the \suzaku operations team  for performing  this observation 
and providing software and calibration for the data analysis. 
This research has also made use of data obtained from the 
High Energy Astrophysics Science 
Archive Research Center (HEASARC), provided by NASA's Goddard Space 
Flight Center.



\bibliographystyle{apj}      
\bibliography{xray_nov_2009}   

\clearpage
\begin{table}
\begin{threeparttable}
\centering
\caption{The Observation Log}
{\footnotesize 
\begin{tabular}{l c c c c}
\hline\hline
Observation & Date & ct s$^{-1}$  & Flux \tnote{1} & Exposure time \\

\hline
\textsl{XMM 1}  &  2001-05-16 &19.27 & $2.31 $ & 39.6 \\
\hline
\textsl{XMM 2}  &  2002-11-22 &3.15  & $0.56 $ & 150.9 \\
\hline
\textsl{Suzaku 1}  &  2005-11-10 &0.45 & $0.87$ & 120   \\
\hline
\textsl{Suzaku 1-i1}  & 2005-11-10 & 0.32 & $0.70$ & 80.0 \\
\textsl{Suzaku  1-i2}  & 2005-11-10 & 0.71 & $1.26$ &  39.6  \\
\hline
\textsl{Suzaku 2}   & 2008-11-06 & 2.06  & $2.42$ &  275  \\
\hline
\textsl{Suzaku 2-i1}   & 2008-11-06 & 1.37 & $1.76$ &   129.2 \\
\textsl{Suzaku 2-i2}  & 2008-11-06 & 2.38 & $2.75$ & 92.9 \\
\textsl{Suzaku 2-i3}  & 2008-11-06  & 3.63 & $3.89$ &  40.6 \\
\hline
\textsl{Suzaku 3}  & 2008-11-23 &  1.42 &$1.79 $ &  78 \\
\hline
\end{tabular}
}
\begin{tablenotes}[para]
\item[] {{\it i1, i2} etc refer to the intensity selected data, see 
text for details} \\
\item[1] {The observed 2-10\,keV flux in units
 $10^{-11} {\rm  ergs\,cm^{-2}\,s^{-1}}$ } \\
\item[2] {Mean exposure time in the CCD instruments in ks} \\
\item[3] {Mean count rate in the 0.5-10\,keV band given per FI XIS in the case of {\it Suzaku} or  
for the pn in the case of {\it XMM} }  \\
    \end{tablenotes}
       \end{threeparttable} 
\end{table}

\begin{table}
\begin{threeparttable}
\centering
\caption{The Spectral Model}
{\footnotesize 
\begin{tabular}{l l l l }
\hline\hline
Parameter & 2005 & 2008a  & 2008b  \\

\hline
$\Gamma$  &  2.47$\pm0.02$ & $l$ &  $l$  \\
$PL_{norm}^1$  & $0.58\pm0.01$ & $1.818\pm0.004$  & $1.283\pm0.005$  \\
$N_{H4} ^2$  &  $2.17\pm 0.23^2$  & $l$ & $l$  \\
Log $\xi_4^3$  & 3.40$\pm0.01$  & $l$  & $l$  \\
$N_{H5} ^2$  & 1.17$\pm0.04$ &  $l$ &  $l$  \\
Log $\xi_5$   & 0.18$^{+0.15}_{-0.16}$ &  $l$  &  $l$  \\
$Covering(N_{H5})$   & 71$\pm 2$\%  &  $35\pm 2$ \% &  $38\pm 2$\% \\
$Pex_{abund}^4$  & 0.40$\pm 0.03$ &  $l$  & $l$  \\
$Pex_{norm}^4$  & $3.84\pm 0.19$ &  $5.41\pm0.26$ &  $4.84\pm0.28$ \\
$N_{H6} ^1$  & $50.0^{p}_{-5.0}$ &  $l$ &  $l$ \\
Log $\xi_6$   & 2.62$\pm 0.05$  &  $l$  &  $l$  \\
Log $Reflion_{\xi}$  & $2.94\pm 0.07$ &  $l$ &  $l$  \\
$Reflion_{norm}$ & $1.03 \times 10^{-7}$  &  $l$ & $l$ \\
\hline
\end{tabular}
}
\begin{tablenotes}[broadfit]
\item[1] {PL normalization in units  $10^{-2} {\rm photon\, cm^{-2}s^{-1}}$ at 1 keV}  \\
\item[2] {Column densities in units of $10^{23} {\rm cm^{-2}}$ }  \\
\item[3] {This zone is outflowing at $\sim 5500$ km/s -see \citealt{lobban10a} and also \citealt{collinge01a,steenbrugge09a}}  \\
 \item[4] {Parameters of the pexrav component, abundance as fraction of solar and 
normalization in units  $10^{-2} {\rm photon\, cm^{-2}s^{-1}}$ at 1 keV; } \\
\item[] {Normalizations are in  ${\rm photons\,cm^{-2}\,s^{-1}}$} \\
\item[] {$l$ indicates a parameter that was linked to be the same for all epochs} \\
\item[] {All components are also fully covered by the Galactic column of neutral gas with 
$N_H(Gal)=1.3 \times 10^{20} {\rm cm^{-3}}$ plus three zones of ionized gas determined from 
HETG feature measurements \citep{lobban10a} and thus frozen in this fit. 
$N_{H1}= 9.6 \times 10^{19} {\rm cm^{-2}}, {\rm log}\, \xi_1=0.90; 
N_{H2}=5.12 \times 10^{20}{\rm cm^{-2}}, {\rm log}\, \xi_2=2.08; 
N_{H3}=1.08 \times 10^{21} {\rm cm^{-2}}, {\rm log}\, \xi_3=2.14$ } \\
\item[] $N_{H5}$ covers a fraction of the powerlaw (only) \\
\item[] $N_{H6}$ covers the ionized reflector  (only) \\ 
   \end{tablenotes}
       \end{threeparttable} 
\end{table}

\clearpage

\begin{table}
\begin{threeparttable}
\centering
\caption{X-ray Line Emission in the 3-10 keV band}
{\footnotesize
\begin{tabular}{l|c c c|c c c|c}
\hline\hline
Observation  & F$_{6.40}$ \tnote{1} & EW$_{6.40}$  & $\Delta\chi^{2}_{6.40}$  & F$_{5.44}$ \tnote{1} & EW$_{5.44}$ & $\Delta\chi^{2}_{5.44}$ & $\chi^{2}/d.o.f.$ \tnote{2} \\
\hline
\textsl{XMM 1} &  $23.79^{+3.28}_{-3.30}$ & $95^{+15}_{-15}$ & 143 & $0.86^{
+3.24}_{-0.86}$ & $23^{+87}_{-23}$ & 0  & 117/109 \\
\hline
\textsl{XMM 2} &  $18.07^{+2.58}_{-2.16}$ & $205^{+16}_{-16}$ & 158 & $2.50^{+
1.91}_{-2.09}$ & $37^{+30}_{-30}$ & 3 & 146/108 \\
\hline
\textsl{Suzaku 1} &  $19.05^{+1.60}_{-1.60}$ & $155^{+24}_{-24}$ & 379 & $5.03^{+1.44}_{-1.45}$ & $46^{+16}_{-16}$ & 32 & 150/100 \\
\hline
\textsl{Suzaku 1-i1}  & $18.30^{+1.86}_{-1.87}$ & $201^{+20}_{-21}$ & 261 & $5.18^{+1.59}_{-1.60}$ & $46^{+14}_{-14}$ & 27 & 107/100 \\
\textsl{Suzaku 1-i2}  & $20.99^{+3.06}_{-3.07}$ & $149^{+22}_{-22}$ & 127 & $4.84^{+2.87}_{-2.87}$ & $26^{+15}_{-15}$ & 8  & 136/100  \\
\hline
\textsl{Suzaku 2}  &  $22.31^{+1.65}_{-1.65}$ & $91^{+7}_{-7}$ & 160 & $4.23^{+1.57}_{-1.57}$ & $13^{+5}_{-5}$ & 19  & 182/100 \\
\hline
\textsl{Suzaku  2-i1} &  $21.95^{+2.02}_{-2.02}$ & $120^{+11}_{-11}$ & 269 & $2.35^{+1.99}_{-2.00}$ & $10^{+8}_{-8}$ & 4  & 135/100 \\
\textsl{Suzaku 2-i2} &  $20.52^{+2.95}_{-2.95}$ & $74^{+11}_{-11}$ & 131 & $5.81^{+2.88}_{-2.88}$ & $16^{+8}_{-8}$ & 11  & 117/100 \\
\textsl{Suzaku 2-i3} &  $27.75^{+5.15}_{-5.15}$ & $75^{+14}_{-14}$ & 79 & $6.80^{+5.10}_{-5.10}$ & $14^{+10}_{-10}$ & 5  & 114/100 \\
\hline
\textsl{Suzaku 3}  & $19.98^{+3.02}_{-3.19}$ & $107^{+14}_{-14}$ & 142 & $4.31^{+2.69}_{-3.05}$ & $17^{+12}_{-12}$ & 19  & 120/100 \\
\hline
\end{tabular}
}

\begin{tablenotes}[para]
\item[] {{\it i1, i2} etc refer to the intensity selected data, see text for details} \\ 
\item[1]{Line normalization in units 10$^{-6}$ photons cm$^{-2} s^{-1}$ for 
 a line having width $\sigma=50$ eV} \\
\item[2] {Fit statistic including both of the lines} \\
\item[] Line equivalent widths are measured against the total observed continuum \\
     \end{tablenotes}
       \end{threeparttable}
\end{table}


\clearpage

\end{document}